\documentclass[twocolumn,showpacs,preprintnumbers,amsmath,amssymb]{revtex4}
\usepackage{graphicx}
\usepackage{dcolumn}
\usepackage{bm}
\usepackage{array}
\usepackage{amssymb}
\usepackage{amssymb,bm,mathrsfs,bbm,amscd}
\usepackage{subfigure}
\usepackage{url}
\usepackage{longtable}
\usepackage{supertabular}

\begin{document}

\title[]{Dynamic Patterns of Academic Forum Activities}

\author{Zhi-Dan Zhao$^{1,2}$},
\author{Ya-Chun Gao$^{3}$}
\author{Shi-Min Cai$^{1,2}$}
\email{shimin.cai81@gmail.com}
\author{Tao Zhou$^{1,4}$}
\email{zhutou@ustc.edu}
\address{$1$ CompleX Lab, Web Sciences Center, University of Electronic Science and Technology of China, Chengdu 611731, P. R. China}
\address{$2$ State Key Laboratory of Networking and Switching Technology (Beijing University of Posts and Telecommunications), Beijing, 100876, P. R. China}
\address{$3$ School of Management and Economics, University of Electronic Science and Technology of China, Chengdu, 611731, P. R. China}
\address{$4$ Big Data Research Center, University of Electronic Science and Technology of China, Chengdu 611731, P. R. China}

\begin{abstract}
A mass of traces of human activities show rich dynamic patterns.
In this article, we comprehensively investigate the dynamic patterns of 50 thousands of
researchers' activities in \emph{Sciencenet}, the largest multi-disciplinary academic
community in China. Through statistical analyses,
we found that (i) there exists a power-law scaling
between the frequency of visits to an academic forum and the number of corresponding visitors,
with the exponent being about $1.33$;
(ii) the expansion process of academic forums
obeys the Heaps' law, namely the number of distinct visited forums to the number of visits grows in a power-law form
with exponent being about 0.54;
(iii) the probability distributions of time intervals
and the number of visits taken to revisit the same academic forum both follow power-laws,
indicating the existence of memory effect in academic forum activities.
On the basis of these empirical results, we propose a dynamic model that incorporates the
exploration, preferential return with memory effect, which can well reproduce
the observed scaling laws.
\end{abstract}


\maketitle

\section{Introduction}
It has been deemed that complicated factors are affecting the dynamic patterns of
human activities, such as the priority of task~\cite{Barabasi2005,Vazquez2005prl,Vazquez2006pre,Anteneodo2009pre}, individual interest~\cite{Han2008,Shang2010,Zhao2013},
memory effects~\cite{Szell2012,Yamasaki2005,Vazquez2007,Goh2008,Cai2009,Zhao2012}, deadline effects~\cite{Alfi2007np}, social contacts~\cite{Wu2010pnas,StehlE2010,Zhaok2011pre,Zhaok2011a}, and so on.
Relevant practical applications range from information
spreading~\cite{Onnela2007,Iribarren2009,lv2011njp,Meloni2011}, decision
making~\cite{Salganik2006} to advertising~\cite{Dukas2004,Reis2006} and
recommendation~\cite{lv2012pr,Zhou2009njp,Guimer2012}.

Although large amounts of empirical results on human dynamics have
been reported in various fields~\cite{Wu2010pnas,Zhou2008epl,Malmgren2009science,Radicchi2009pre,Zhao2012physa}, researchers' activities in academic forums
are rarely investigated and still not clearly
understood as they usually accompanies with
many endogenous and exogenous factors, including the individual preference and professional background,
the quality of a forum, the content of a post, and so on.
To fill this gap, we study a data set sampled from
\emph{Sciencenet} (http://www.sciencenet.cn/) that contains
the academic forum activities of many Chinese researchers.
We observe novel dynamic patterns characterized
by the following statistical features: (i) the power-law relation between the frequency of visits to an academic
forum and the number of corresponding visitors; (ii) the Heaps' law~\cite{Heaps1978} in the expansion process; (iii) the memory effect. We further propose a dynamic model that well reproduces the empirical observations.

\section{Empirical Results}
Sciencenet is the largest multi-disciplinary academic community in China,
which contains a blog system, a bulletin board system (BBS, consisting of $60$ academic forums),
and a virtual social network of researchers. Our data set keeps track of activities in the BBS
between October/1/2007 and July/7/2011, composed of 366,524 records from
49,578 researchers. Each record includes researcher ID, academic forum ID, posting/reviewing topic ID,
and timestamp with resolution of minute. Table~\ref{tab1} presents the names, the visiting frequencies (i.e, total visits
from all researchers), and the number of visitors of the 60 academic forums, ranked in the descending order of visiting frequencies.


\begin{longtable}[!hbp]{|c|p{4cm}|c|c|}
\hline
\hline
Rank & Forum Name & Frequency & \#visitors  \\
\hline
1 & Mathematics & 28,218 & 5,849  \\
\hline
2 & Materials Science & 23,610 & 6,293  \\
\hline
3 & Geology and Geophysics & 21,363 & 4,025  \\
\hline
4 & Oceanography & 21,265 & 4,019  \\
\hline
5 & Reading & 18,586 & 5,315 \\
\hline
6 & Research Experience & 15,757 & 6,713 \\
\hline
7 & Tea Break & 13,351 & 4,209 \\
\hline
8 & Fund Application & 11,672 & 4,045 \\
\hline
9 & Article & 11,511 & 4,983  \\
\hline
10 & Analytical Chemistry & 9,228 & 2,769  \\
\hline
11 & High-polymer Chemical & 8,271 & 2,536  \\
\hline
12 & Nanotechnology & 8,211 & 2,815  \\
\hline
13 & Amorphous Alloy and Glass & 8,148 & 1,151  \\
\hline
14 & Computer Science & 7,834 & 2,709 \\
\hline
15 & Chemical Engineering & 7,541 & 2,208  \\
\hline
16 & Catalysts & 7,541 & 2,208  \\
\hline
17 & Organic & 6,952 & 2,210  \\
\hline
18 & Physics & 6,702 & 2,547  \\
\hline
19 & Chemistry & 6,653 & 2,860 \\
\hline
20 & Physical Resources & 6,289 & 2,402 \\
\hline
21 & References & 6,102 & 3,115 \\
\hline
22 & Study Abroad & 6,085 & 2,781 \\
\hline
23 & Optics & 5,984 & 1,759 \\
\hline
24 & Molecular and Cellular Immunology & 5,544 & 2,132 \\
\hline
25 & Electronic Power & 5,539 & 1,841  \\
\hline
26 & Mechanics & 5,306 & 1,736  \\
\hline
27 & Control of Intelligent Modeling & 5,044 & 1,763  \\
\hline
28 & Electrochemistry & 4,961 & 1,453  \\
\hline
29 & Crystal & 4,775 & 1,881  \\
\hline
30 & Mechanical & 4,772 & 1,483  \\
\hline
31 & Theoretical Physics & 4,732 & 1,804  \\
\hline
32 & Communication & 4,691 & 1,600  \\
\hline
33 & Civil Engineering & 4,164 & 905 \\
\hline
34 & Botany and Zoology & 3,973 & 1,577 \\
\hline
35 & Clinical Medicine & 3,149 & 1,021  \\
\hline
36 & Condensed Matter & 3,130 & 1,195  \\
\hline
37 & Examination & 2,849 & 1,460  \\
\hline
38 & Resources and Environment & 2,648 & 1,271  \\
\hline
39 & Fine Chemicals & 2,464 & 772  \\
\hline
40 & Materials and Methods & 2,452 & 1,397  \\
\hline
41 & Pharmaceutical Chemistry & 2,357 & 677  \\
\hline
42 & Chinese Medicine & 2,170 & 548  \\
\hline
43 & History of Science & 2,142 & 991  \\
\hline
44 & Engineering & 2,040 & 1,080  \\
\hline
45 & Management & 1,920 & 978 \\
\hline
46 & Academic Exhibition & 1,920 & 978 \\
\hline
47 & GPS/RS/GIS & 1,862 & 757  \\
\hline
48 & Genetics & 1,776 & 769  \\
\hline
49 & Agronomy & 1,776 & 769  \\
\hline
50 & General Chemistry & 1,692 & 824  \\
\hline
51 & Astrophysics & 1,565 & 509 \\
\hline
52 & Biophysical and Bioinformatics & 1,485 & 711 \\
\hline
53 & Molecular Simulation & 1312 & 571  \\
\hline
54 & Energy Science & 1,233 & 546  \\
\hline
55 & Brain Science & 1,161 & 553  \\
\hline
56 & Petroleum Exploration & 1,089 & 450  \\
\hline
57 & Others & 631 & 500  \\
\hline
58 & Cryptography & 625 & 251  \\
\hline
59 & Alloy & 427 & 176  \\
\hline
60 & Medical and Health Management & 275 & 196  \\
\hline
\hline
\caption{\label{tab1} Basic information of the 60 academic forums.}
\end{longtable}


\begin{figure}[htp]
\begin{center}
\includegraphics[width=0.45\textwidth]{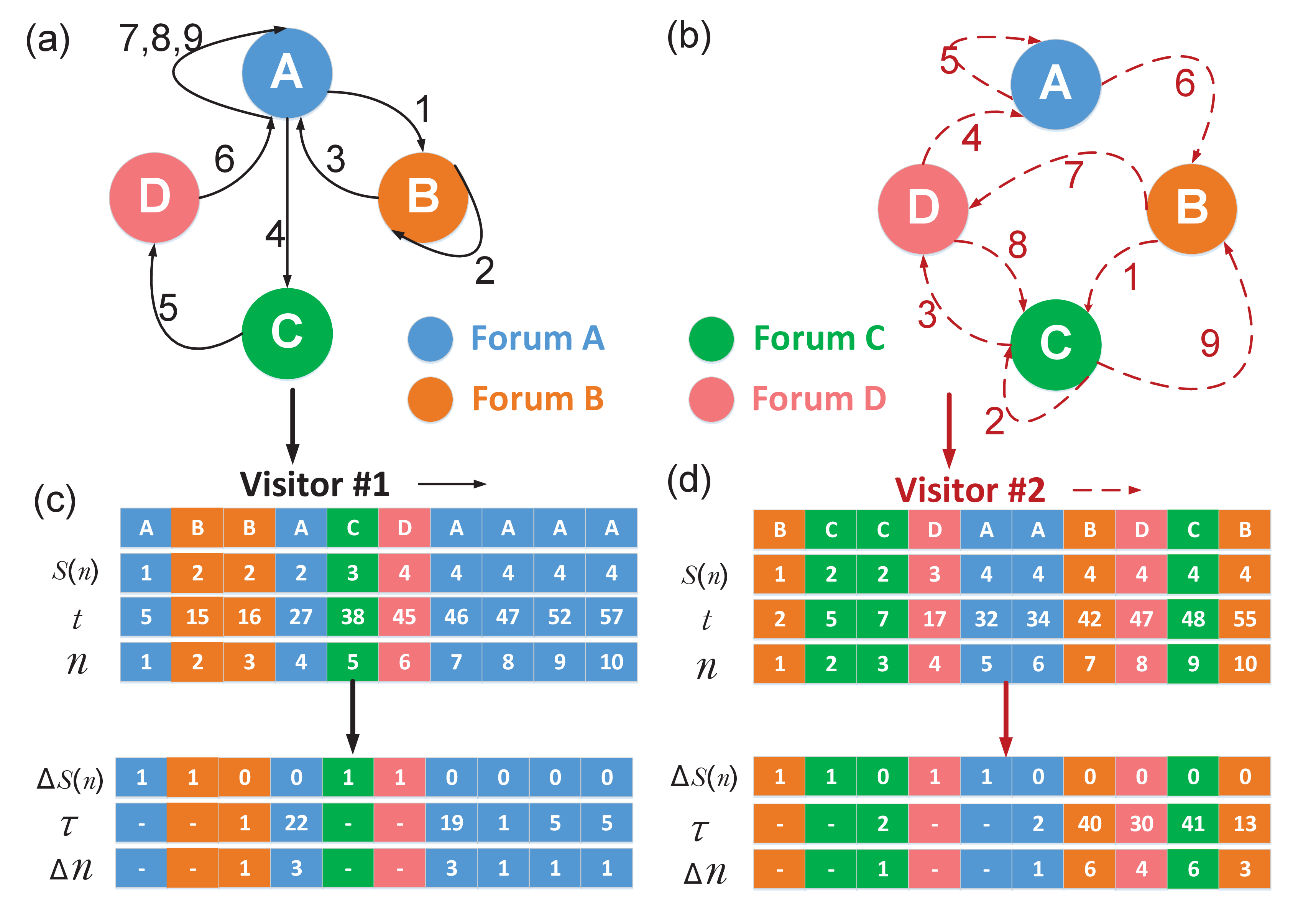}
\caption{\label{fig0} (Color online) An illustration of concepts in view of two visitors' activities.
(a)-(b) The transition process among different academic forums (black solid and red dash lines represent
visitor $\#1$ and visitor $\#2$, respectively).
(c)-(d) The illustration of visiting sequences, as well as other relevant parameters of the two visitors, including the number of distinct academic forums $S(n)$, the real time $t$,
the number of visits $n$, the exploration of new academic forums $\Delta S(n)$, the real time interval
$\tau$ and the click time interval $\Delta n$.}
\end{center}
\end{figure}

\begin{figure}[htp]
\begin{center}
\includegraphics[width=0.45\textwidth]{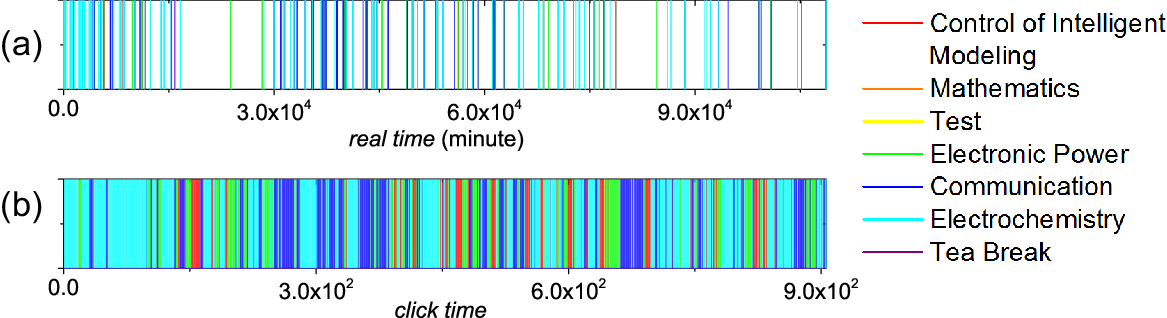}
\caption{\label{fig1} (Color online) A typical visitor's transition process on academic forums along with (a) real time and (b) click time,
suggesting a heterogeneous dynamic pattern that commonly exists in other online human dynamics~\cite{Zhao2012physa}. The colored vertical line denotes
the different academic forums.}
\end{center}
\end{figure}

At the aggregated level, as shown in Table~\ref{tab1}, both the number of visits and
the number of visitors are heterogeneously distributed among forums. While at the individual level, the visiting behavior is also heterogeneous,
indicated by the burstiness that a visitor usually stays in a forum for long time and then glances over several forums.
Figure~\ref{fig0} illustrates the transition process of two example visitors, with visiting sequences shown in the first
rows of Fig.~\ref{fig0}(c) and \ref{fig0}(d). Figure~\ref{fig1} shows the transition process of a typical real visitor,
with two different scales: (a) real time and (b) click time. The colored vertical lines represent different academic forums.

\begin{figure}[htp]
\begin{center}
\includegraphics[width=0.45\textwidth]{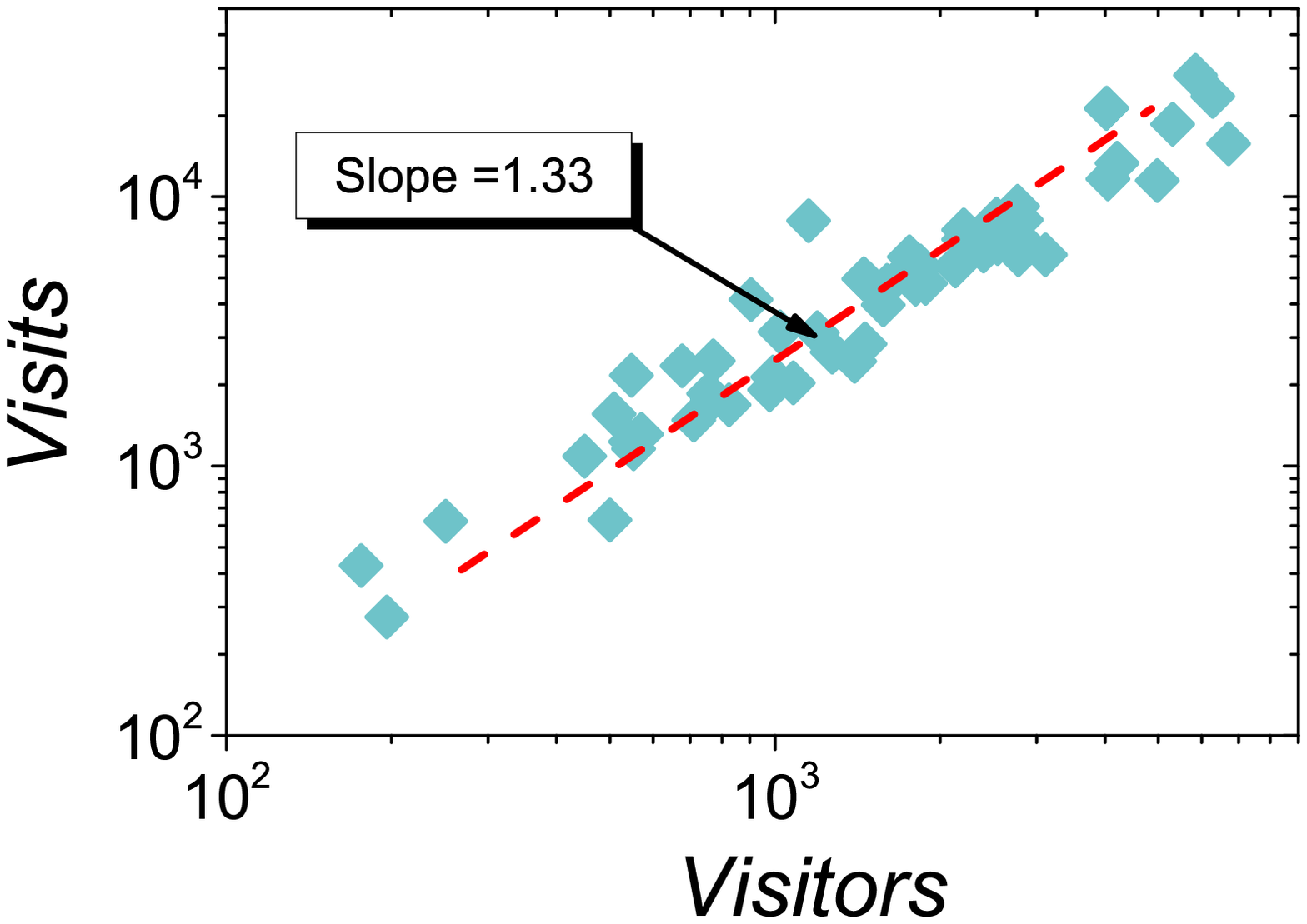}
\caption{\label{fig2} (Color online) A power-law relation between the frequency of visits to
an academic forum and the number of corresponding visitors. Each data point stands for an academic forum.}
\end{center}
\end{figure}

As shown in Fig.~\ref{fig2}, there is a superlinear correlation between the number of visits
to a academic forum (denoted by $F$) and the number of corresponding visitors (denoted by $P$),
as $F \sim {P^\gamma }$ with $\gamma \approx 1.33$. This superlinear relationship maybe resulted from
two reasons. Firstly, the forum that attracts more visitors is usually of higher quality, namely a post in
this forum can attract a higher fraction of visitors in average.
Secondly, many visits are induced by some previously comments and replies~\cite{Zha2015}, in particular, visitors often
care much about the replies to their own posts and comments. Such social cascading process may lead to a
superlinear growing trend since the maximum possible volume of social interaction is of the order $P^2$.


\begin{figure}[htp]
\begin{center}
\includegraphics[width=0.45\textwidth]{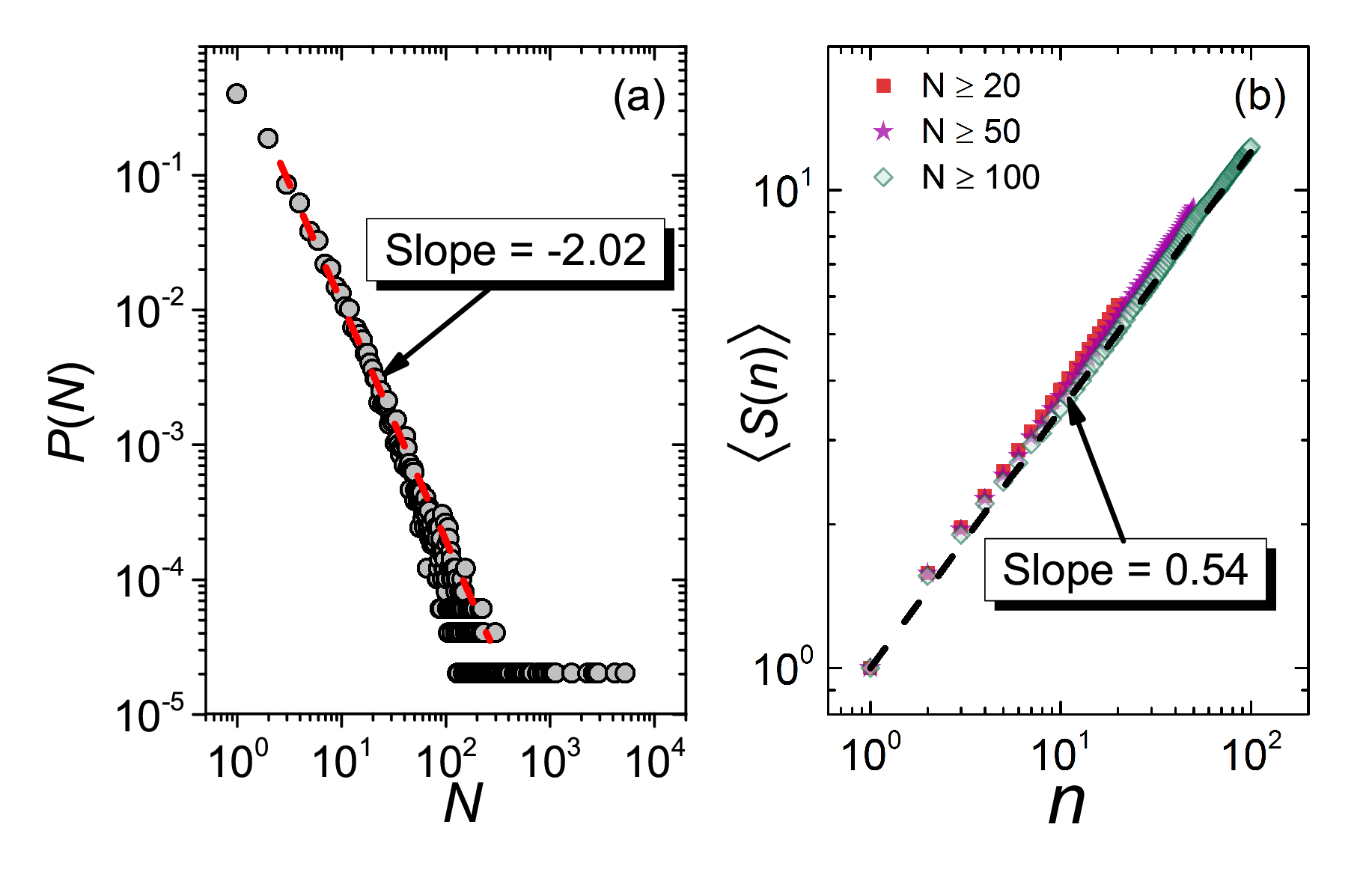}
\caption{\label{fig4} (Color online) (a) The probability
distribution of the number of visits $N$ for all visitors. (b) The scaling behavior in the expansion process. The black dash line
indicates a power-law, $S \sim {n}^\alpha$, with scaling exponent $\alpha  \approx 0.54$. The red squares, purple stars and green diamonds
represent three groups of visitors with more than $20$, $50$ and $100$ visits,
respectively. }
\end{center}
\end{figure}

The expansion process of a visitor can be described by the number of distinct visited forums $S(n)$,
where $n$ is the number of visits of this visitor.
Taking visitor $\#1$ in Fig.~\ref{fig0} as an example, his visiting sequence is
$\{$A$, $B$, $B$, $A$, $C$, $D$, $A$, $A$, $A$, $A$\}$,
and the corresponding $S(n)$ sequence is $\{1, 2, 2, 2, 3, 4, 4, 4, 4, 4\}$, as shown in the second row of Fig.~\ref{fig0}(c).
The activity level is a key feature of an user~\cite{Zhao2012physa}, which may affect the expansion process. Here we use the total number of
visits $N_i$ to quantify the activity level of user $u_i$. As shown in Fig.~\ref{fig4}(a), the distribution of users' activity levels is very heterogeneous, following
a power law with exponent about $2.02$, in accordance with the empirical results in other online activities~\cite{Zhao2012physa}.
In common sense, users with different activity levels may behave differently,
hence we pick out three groups of visitors whose visits
are more than $20$, $50$ and $100$ times, respectively. Their expansion sequence $S(n)$ in respect
to the number of visits $n$ is presented in Fig.~\ref{fig4}(b), suggesting a robust scaling behavior
$S \sim n^{\alpha}$, with the power-law exponent $\alpha \approx 0.54$, similar to some observations in
human mobility~\cite{Song2010,Chmiel2009}.

\begin{figure}[htp]
\begin{center}
\includegraphics[width=0.45\textwidth]{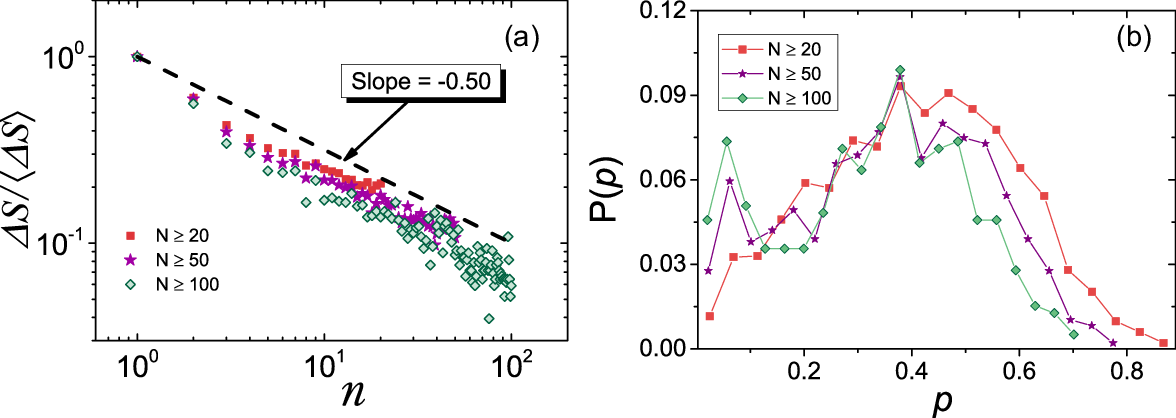}
\caption{\label{fig5} (Color online) (a) The scaling behavior of
rescaled exploration $\frac{\Delta S(n)}{\langle\Delta S(n) \rangle}$
in respect to the number of visits $n$.
(b) The probability distributions of the values of $p$ for individual visitors. The definition
of groups is same to that of figure~\ref{fig4}.}
\end{center}
\end{figure}

To see the preference to visit a new forum, we look at the exploration sequence $\Delta S(n) = S(n+1)-S(n)$.
Simple examples about the calculation of $\Delta S(n)$ are shown in Fig.~\ref{fig0}(c) and \ref{fig0}(d). As shown in
Fig.~\ref{fig5}(a), the rescaled exploration $\frac{\Delta S(n)}{\langle \Delta S(n) \rangle}$ scales in a power-law
form with the number of visits, as $\frac{\Delta S(n)}{\langle \Delta S(n) \rangle} \sim n^{-\beta}$, where $\beta \approx 0.50$
and the rescaling factor $\langle \Delta S(n) \rangle$ is averaged over all visitors in the considered group.
The result confirms the scaling behavior found in expansion process since $\alpha + \beta \approx 1$.
Given the target visitor, if we denote $P_{new}$ the probability to visit a new forum and $1-P_{new}$
the probability to return to a previously visited forum, then the above result suggests a scaling relation
$P_{new} \sim  p n^{-\beta}$, where $p$ is a constant. Figure~\ref{fig5}(b) presents the
distributions of $p$ for the three groups of visitors, which are bell-shaped curves with mean values
as $\langle p \rangle \approx 0.41$,
$\langle p \rangle  \approx 0.36$ and $\langle p \rangle \approx 0.32$ for $N\geq 20$, $N\geq 50$ and $N\geq 100$, respectively.

\begin{figure}[htp]
\begin{center}
\includegraphics[width=0.45\textwidth]{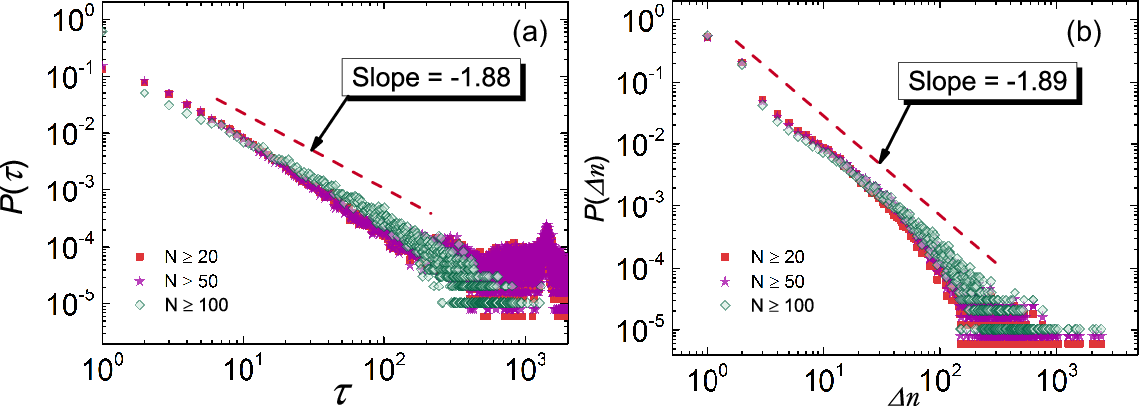}
\caption{\label{fig6} (Color online) The memory effect in visiting activities,
suggested by the power-law distributions of real time interval ${\tau}$ and
(b) the number of visits $\Delta {n}$. The guiding curve follows a power-law
form, ${\tau}^{-\gamma}$ and $\Delta {n}^{ - \xi}$ with $\gamma \approx 1.88$ and $\xi \approx1.89$,
obtained by the maximum likelihood method~\cite{Clauset2009}.
The definition of groups is same to that of figure~\ref{fig4}.}
\end{center}
\end{figure}

Memory is a very significant feature of human activities~\cite{Han2008,Zhao2013,Szell2012,Yamasaki2005,Vazquez2007,Goh2008,Cai2009,Zhao2012,Zhao2012physa,Guo2015},
which can be characterized by the probability distribution of the real time interval $\tau$ between two consecutive visits to the same forum or the
number of visits $\Delta n$ taken to revisit the same forum.
For example, $\Delta n$ for the forum $B$ in the
visiting sequence of visitor $\#2$ are $6$ and $3$, as
shown in Fig.~\ref{fig0}(d). More examples for both $\tau$ and $\Delta n$
can be found in Fig.~\ref{fig0}(c) and \ref{fig0}(d).
Figure~\ref{fig6} reports the probability
distributions of ${\tau}$ and $\Delta n$ of the three
groups of visitors. Both $P(\tau)$ and $P(\Delta n)$ can be well fitted by power laws with
exponents being $1.88$ and $1.89$, respectively. Such distributions suggest the existence of the memory effect
since a visitor has higher probability to return to recently visited forums.

\section{Model}
The above-mentioned empirical results provide us insights to
the dynamic patterns of visiting to academic forums.
In particular, the memory effect assigns higher probability to a visitor to return to
the forums being visited recently.  Accordingly, we propose a model
incorporating three generic ingredients: (i) the
exploration, indicating the tendency to visit a new
forum; (ii) the preferential return, indicating the tendency to visit a previously visited forum,
with frequently visited ones being more attractive;
(iii) the memory effect, indicating the tendency to revisit a recently visited forum.
Notice that, the preferential return can be mathematically considered as a kind of memory
with infinite memory length, however, we distinguish it from memory effect since the preferential return may be resulted
from the quality of a forum and the match between the forum's content and the
visitor's professional background.

\begin{figure}[htp]
\begin{center}
\includegraphics[width=0.45\textwidth]{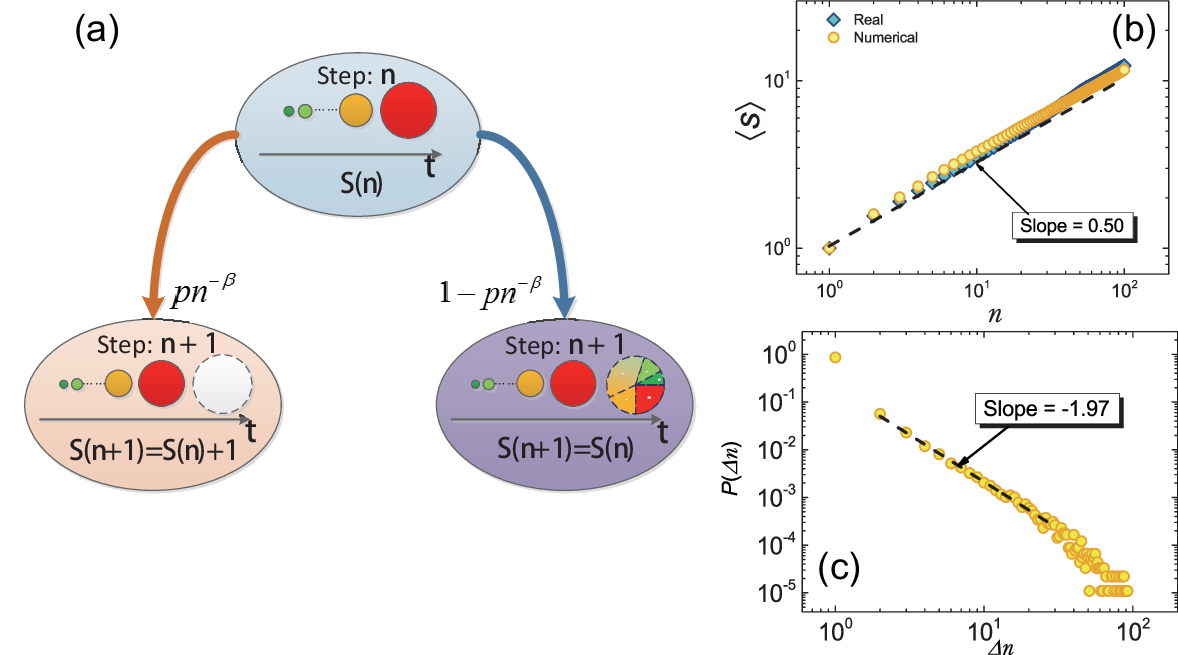}
\caption{\label{fig7} (Color online) (a) The schematic of the individual dynamic model.
(b) The scaling behaviors of expansion $S(n)$ in respect to $n$ for empirical data and the model.
(c) The power-law distribution of return interval $\Delta {n}$
obtained from the model.
The parameters are set as $N = 100$, $\langle p \rangle = 0.41 $, $\beta = 0.50 $, $\xi = 1.89$,
and the number of visitor is $10^{3}$, namely all the results can be considered as being averaged over $10^3$ independent runs.}
\end{center}
\end{figure}

Figure~\ref{fig7}(a) illustrates a
schematic for the individual dynamic model. More specifically, after the $n$th visit, the target visitor has two choices for the next visit: (i)
to visit a new academic forum with probability
$P_{new}$ that depends on the number of previous visits $n$; (ii) to revisit one of the $S(n)$ previously
visited forums with the complementary probability $(1-P_{new})$.
If the visitor visits
a new forum at the $(n+1)$th visit, the expansion sequence updates as $S(n+1)=S(n)+1$,
otherwise it keeps $S(n+1)=S(n)$. Initially, $S(1)=1$, and the probability $P_{new}$
evolves as
\begin{equation} \label{eq2}
{P_{new}} \propto \frac{{dS}}{{dn}} = p{n^{ - \beta }}.
\end{equation}
In our model, the parameter $\beta = 0.50$ is obtained from the empirical
scaling exponent for the case $N \ge 20$ (see Fig.~\ref{fig5}(a)), while the
parameter $\langle p \rangle =0.41$ is estimated from the mean value of
$p$ for the case $N \ge 20$ (see Fig.~\ref{fig5}(b)).


When considering the return to a visited forum, we combine preferential return and memory effect into a unified probability. The preferential return indicates that every visit to a forum will contribute to the probability to revisit this forum, while the memory effect suggests a power-law decay of the contribution strength. Given a forum $i$, if it appears $f_i$ times in the target visitor's visiting sequence, we denote the discrete time interval between the $j$th appearance to the current time step $(n+1)$ as
\begin{equation}
\Delta n_j^{(i)}=n+1-T_j^{(i)},
\end{equation}
where $T_j^{(i)}$ is the very time step when $i$ just appears $j$ times. As discussed above, every visit will be counted but the contribution decays with the length of time interval, so the probability to visit the forum $i$ is
\begin{equation}
\Pi_{i} = \frac{\sum_{j=1}^{f_i} \left(\Delta n_{j}^{(i)}\right)^{- \xi}}{\sum\limits_{h \in {\Gamma _u}} \sum_{j=1}^{f_h} \left(\Delta n_{j}^{(i)}\right)^{- \xi}},
\label{prob}
\end{equation}
where $\Gamma_u$ is the set of forums being visited by the target user $u$, and the decaying factor $\xi \approx 1.89$ is estimated from Fig.~\ref{fig6}(b).
The validation of the individual dynamic model is demonstrated in Fig.~\ref{fig7}(b) and ~\ref{fig7}(c), from which one can see that
the numerical results consist well with the empirical scaling
behaviors for both expansion and revisit.

\section{Conclusion}
In this paper, we comprehensively investigate the dynamic patterns of researchers' academic
forum activities in Sciencenet. We show a power-law scaling between
the frequency of visits to an academic
forums and the number of corresponding visitors, which is similar to the allometric scaling law
found in biology systems~\cite{Kleiber1932,West1997}.
Meanwhile, at the individual level, the number of distinct visited forums $S(n)$
increases with the number of visits $n$ in a power-law behavior, obeying the well-known Heaps' Law~\cite{Heaps1,Heaps2}, which is also
similar to the previous studies on
portal browsing activity and human mobility~\cite{Song2010,Chmiel2009}. The memory effect
in academic forum activity is unveiled by the power-law
distributions of the real time interval ${\tau}$ and the
number of visits $\Delta {n}$ taken to revisit the same academic forum.

Inspired by these empirical results and a previous theoretical model~\cite{Song2010}, we propose a dynamic model, incorporating with the
exploration of new academic forums and the preferential return with
memory effect. Through extensively experimental testing, the numerical results from this dynamic model agree well with the empirical observations,
suggesting its validation. We have also checked that the lack of each of the three ingredients will lead to huge deviation from the real statistics.

\section*{Acknowledgments}
We acknowledge Xiaoyan Yuan for sharing the experimental data.
This work is partially supported by the National Natural Science Foundation of China (Grant Nos. 11222543 and 61433014),
Special Project of Sichuan Youth Science and Technology Innovation Research
Team (Grant No. 2013TD0006), and the Open Foundation of State Key Laboratory of
Networking and Switching Technology (Beijing University of Posts and Telecommunications) (Grant No. SKLNST-2013-1-18).

%

\end{document}